# Increasing efficiency and reducing bias when assessing HPV vaccination efficacy by using non-targeted HPV strains

Lola Etievant[1]*, Joshua N. Sampson[1], Mitchell H. Gail[1]*

lola.etievant@nih.gov

gailm@mail.nih.gov

sampsonjn@mail.nih.gov

[1] National Cancer Institute, Division of Cancer Epidemiology and Genetics, 9609 Medical Center Drive, Rockville, MD 20850-9780.

*Corresponding authors

**Abstract**

Studies of vaccine efficacy often record both the incidence of vaccine-targeted virus strains (primary outcome) and the incidence of non-targeted strains (secondary outcome). However, standard estimates of vaccine efficacy on targeted strains ignore the data on non-targeted strains. Assuming non-targeted strains are unaffected by vaccination, we regard the secondary outcome as a negative control outcome and show how using such data can (*i*) increase the precision of the estimated vaccine efficacy against targeted strains in randomized trials, and (*ii*) reduce confounding bias of that same estimate in observational studies. For objective (*i*), we augment the primary outcome estimating equation with a function of the secondary outcome that is unbiased for zero. For objective (*ii*), we jointly estimate the treatment effects on the primary and secondary outcomes. If the bias induced by the unmeasured confounders is similar for both types of outcomes, as is plausible for factors that influence the general risk of infection, then we can use the estimated efficacy against the secondary outcomes to remove the bias from estimated efficacy against the primary outcome. We demonstrate the utility of these approaches in studies of HPV vaccines that only target a few highly carcinogenic strains. In this example, using non-targeted strains increased precision in randomized trials modestly but reduced bias in observational studies substantially.

## 1. INTRODUCTION

There are more than 100 types of the human papillomavirus (HPV) and at least 14 are known to cause cervical cancer. The Food and Drug Administration, as well as other regulatory agencies, have approved HPV vaccines that target and protect against the two most carcinogenic types, HPV types 16 and 18. Although they can also protect against some other types (e.g., HPV 6, 11, 31, 33, 45, 52, and 58), they are documented to have small or negligible effects on most of the non-targeted, non-carcinogenic types (Supplementary Table 1 in Tota *et al.*, 2020). Currently, these vaccines are administered as a two or three dose series depending on the age group. However, in order to simplify administration and encourage uptake in low and middle income countries, there are now both randomized trials and observational studies being conducted to show that a single dose of these HPV vaccines can protect against HPV 16 and/or 18 infection, with the composite endpoint denoted by HPV 16/18 (International Vaccine Institute, 2020; National Cancer Institute (NCI), 2020). Importantly, these studies will also record infections by the non-targeted HPV types. In this paper, our goal is to show how to use the information on these non-targeted HPV types to: (*i*) increase the precision for the estimator of vaccine efficacy on targeted types in randomized trials; and (*ii*) reduce the bias from unmeasured confounders in observational studies. We emphasize that our ability to achieve our goals relies on the assumption that vaccination does not affect non-targeted HPV types.

We first consider a randomized clinical trial where participants are randomized to receive either one dose of an HPV vaccine or a placebo and are then followed to see who develops an HPV 16/18 infection. In order to increase the efficiency of the estimated treatment effect, analyses may adjust for baseline covariates. Relevant to our discussion, Tsiatis *et al.* (2008) and Zhang, Tsiatis and Davidian (2008) proposed a robust semi-parametric approach for handling covariates that augments the standard estimating equations by adding a well-chosen function of those covariates that is unbiased for zero and correlated with the outcome. However, in HPV trials, key covariates, such as sexual behavior may be poorly measured. Therefore, we propose adjusting for a post-randomization event that is highly correlated with the key baseline covariates. Specifically, we modify the method of Zhang *et al.* (2008) and augment our estimating equations by adding a function of the incidence of non-targeted HPV infections.

We next consider observational studies that evaluate the efficacy of an HPV vaccine. For example, in the India Study (Sankaranarayanan *et al.*, 2016), the incidence of HPV 16/18 infection among girls who were vaccinated as part of a clinical trial is compared to the incidence in a survey of unvaccinated girls. In such studies, estimates of vaccine efficacy may be biased if confounders are unmeasured or their effects are not modeled correctly (Lawlor *et al.*, 2004; Fewell, Davey Smith and Sterne, 2007). We can reduce such bias by leveraging our knowledge about infections by non-targeted HPV types, which can be considered to be “negative control outcomes” (Lipsitch, Tchetgen Tchetgen and Cohen, 2010; Shi, Miao and Tchetgen, 2020) or outcomes known to be unaffected by treatment (i.e., vaccination in this paper). We can estimate the effect of the vaccine on the non-targeted types and “substract off ” this effect, which should be null in the absence of confounding, from the effect on the targeted HPV types. We later present assumptions sufficient to assure that this non-null effect can be used to remove (or markedly reduce) bias from unmeasured confounders. One key assumption is that the unmeasured covariates, like sexual behavior, affect the risk of targeted and untargeted types proportionally.

In SECTION 2, we describe the problem formally, introduce notation, and present methods for using the non-targeted types to increase the precision of the estimated vaccine efficacy in randomized trials and for reducing the bias of the estimated vaccine efficacy in observational studies. In SECTION 3, we evaluate the performances of the proposed approaches in simulation studies based on HPV data. In SECTION 4, we apply these methods to a randomized trial (Costa Rica Vaccine Trial, Herrero *et al.*, 2008) and an observational study in India (Sankaranarayanan *et al.*, 2018). We present concluding remarks in SECTION 5. Most technical derivations and details are presented in Web Appendices.

## 2. METHODS

### 2.1. General setting and notation

Our initial goal is to estimate the effect of an HPV vaccine (i.e., treatment) on the risk of having an HPV 16 infection (i.e., primary outcome). We discuss the composite HPV 16 and/or HPV 18 outcome in SECTION 2.4. Our notation includes: $n$, the number of subjects in the study; $T_i$, a binary indicator of vaccination for subject $i$, $i \in \{1, \dots, n\}$; $Y_{1i}$, a binary indicator of an HPV 16 infection for subject $i$, $i \in \{1, \dots, n\}$; $\mathrm{N_{NT}}$, the number of non-targeted HPV types; $Y_{2i}^{(j)}$, a binary indicator of an infection with the $j^{\text{th}}$ non-targeted type for subject $i$, $j \in \{1, \dots, \mathrm{N_{NT}}\}$, $i \in \{1, \dots, n\}$; $Y_{2i}$, the total number of non-targeted infections (i.e., $Y_{2i} = \sum_{j \in [\![1, \mathrm{N_{NT}}]\!]} Y_{2i}^{(j)}$) for subject $i$, $i \in \{1, \dots, n\}$; $\boldsymbol{W}_i$, a set of observed covariates for subject $i$, $i \in \{1, \dots, n\}$; and $\boldsymbol{A}_i$, a set of unobserved covariates for subject $i$, $i \in \{1, \dots, n\}$.

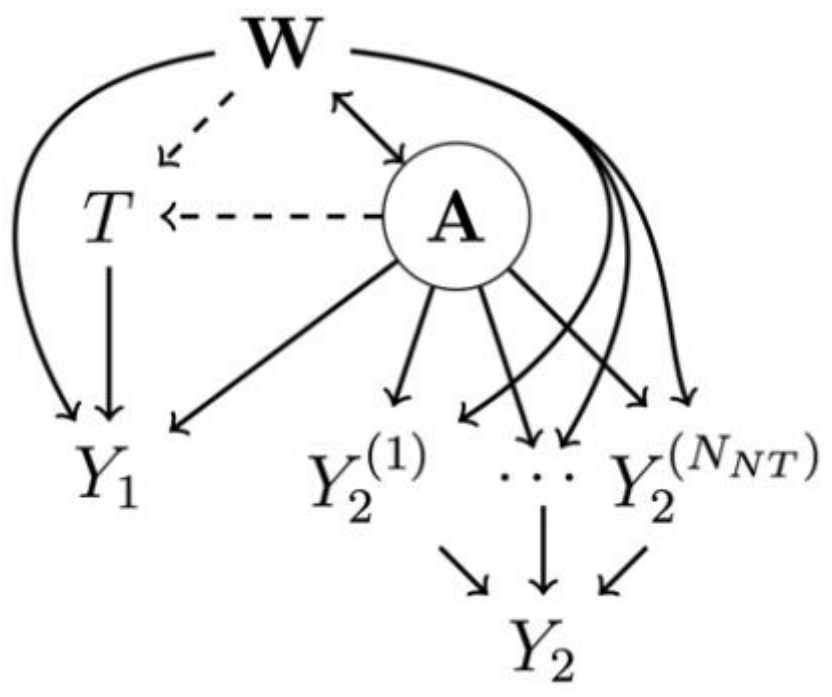

**FIGURE 1-** Graph depicting the relationships among the variables, in both randomized and non-randomized settings. Circled variables denote unmeasured variables. Dashed arrows would be removed in a randomized trial where treatment assignment is independent of measured and unmeasured covariates. The double-headed arrow between $\boldsymbol{W}$ and $\boldsymbol{A}$ encompasses any situation where components of $\boldsymbol{W}$ may causally affect components of $\boldsymbol{A}$ and components of $\boldsymbol{A}$ may causally affect components of $\boldsymbol{W}$, but all affects are uni-directional.

The relationships among the variables are described by the graph in **FIGURE 1**. Regression analyses for subject $i$, $i \in \{1, \dots, n\}$, are based on:

$$\mathrm{E}(Y_{1i} \mid T_i, \boldsymbol{W}_i, \boldsymbol{A}_i) = g(\boldsymbol{A}_i) \exp\{\mu_1 + \beta_1 T_i + q(\boldsymbol{W}_i)\}, \tag{1}$$

$$\mathrm{E}(Y_{2i} \mid T_i, \boldsymbol{W}_i, \boldsymbol{A}_i) = g(\boldsymbol{A}_i) \exp\{\mu_2 + s(\boldsymbol{W}_i)\}. \tag{2}$$

Importantly, $T_i$ does not affect $Y_{2i}$ (i.e., the treatment does not affect the non-targeted types) in equation (2). Note that the same positive and possibly un-known function $g$ appears in equations (1) and (2). In other words, the effect of the unmeasured covariates is assumed to be similar for all HPV types. If a certain behavior increases a woman's risk of an HPV 16 infection by 20%, it is likely to increase her risk of all non-targeted infections by 20%. However, if $g$ in equation (2) is replaced by $h = k \times g$ for some $k > 0$, inference on the treatment effect, $\beta_1$, is unaffected, because the factor $k$ is absorbed into the intercept $\mu_2^* = \mu_2 + \log(k)$. Therefore, we say that the effects of $\boldsymbol{A}$ on targeted and non-targeted types need only be proportional. We can think of $g(\boldsymbol{A})$ as describing the shape, but not the magnitude, of the of the effect of $\boldsymbol{A}$ on non-targeted outcomes. The shape of the effect is the same for targeted and non-targeted outcomes. If $\boldsymbol{W}_i$ is a vector of stratum indicators, we can further weaken the assumptions implied by equations (1) and (2); see SECTION 2.3 and **TABLE 1**, which describes the assumptions needed for the analysis of observational data. In randomized trials, $T_i$ is independent of $\boldsymbol{A}_i$ and $\boldsymbol{W}_i$, and simpler regressions can be used (see SECTION 2.2).

### 2.2. Increasing efficiency in a randomized trial

We first estimate the log relative risk for treatment, $\beta_1$, in a randomized study. From equation (1) and the fact that $T_i$ is independent of $\boldsymbol{A}_i$ and $\boldsymbol{W}_i$, $i \in \{1, \dots, n\}$,

$$E(Y_{1i} \mid T_i) = \exp(\tilde{\mu}_1 + \beta_1 T_i), \tag{3}$$

where $\tilde{\mu}_1 = \mu_1 + \log\{E(\exp[q(\boldsymbol{W}_i) + \log\{g(\boldsymbol{A}_i)\}])\}$. The log-likelihood from equation (3) is

$$\sum_{i=1}^{n}[Y_{1i} \log\{\exp(\tilde{\mu}_1 + \beta_1 T_i)\} + (1 - Y_{1i}) \log\{1 - \exp(\tilde{\mu}_1 + \beta_1 T_i)\}].$$

Traditionally, we could estimate $\boldsymbol{\theta}_1 \equiv (\tilde{\mu}_1, \beta_1)$ by solving the estimating equation $\sum_{i=1}^{n} \boldsymbol{U}_1 (Y_{1i}, T_i; \boldsymbol{\theta}_1) = 0$, where

$$\boldsymbol{U}_1(Y_1, T; \boldsymbol{\theta}_1) = \begin{pmatrix} \frac{Y_1 - p_1}{1 - p_1} \\ \frac{T(Y_1 - p_1)}{1 - p_1} \end{pmatrix}, \tag{4}$$

and $p_1 = \exp(\tilde{\mu}_1 + \beta_1 T)$. We refer to this estimation method as the "Unaugmented" approach (UnAug). This unbiased analysis uses only data on the treatment and primary outcome and ignores other variables.

One way to gain precision is to incorporate baseline covariates into the regression model in equation (3). Alternatively, Tsiatis *et al.* (2008) and Zhang *et al.* (2008) proposed augmenting the estimating equation with a function of baseline covariates. We adapted this approach by augmenting the estimating equation above with a function of the secondary outcome, which, like baseline covariates, is assumed to be independent of treatment. The secondary outcome serves as a surrogate for unmeasured or poorly measured covariates, such as sexual activity, that affect both the primary and secondary outcomes. Let $\boldsymbol{a}_t(Y_2)$ be an appropriate function of the secondary outcome. The augmented estimating equations are of the form $\sum_{i=1}^{n} \boldsymbol{U}_1^* (Y_{1i}, Y_{2i}, T_i; \boldsymbol{\theta}_1) = 0$, with

$$\boldsymbol{U}_1^*(Y_1, Y_2, T; \boldsymbol{\theta}_1) = \boldsymbol{U}_1(Y_1, T; \boldsymbol{\theta}_1) - \sum_{t \in \{0,1\}} \{I(T = t) - \pi_t\} \boldsymbol{a}_t(Y_2),$$

where $\pi_t = P(T = t), t \in \{0,1\}$, and $I(\,)$ is the Indicator function. Because $T$ is independent of $Y_2$, the expected value of the augmenting term is zero, and the solutions to the augmented estimating equation are also un-biased estimates of $\boldsymbol{\theta}_1$. From Zhang *et al.*, (2008), the optimal choice $\boldsymbol{a}_t(Y_2) = \mathrm{E}\{\boldsymbol{U}_1(Y_1, T; \boldsymbol{\theta}_1) \mid Y_2, T = t\}$, $t \in \{0,1\}$, minimizes the variance of $\widehat{\boldsymbol{\theta}}_1$ via the optimal estimating function

$$\boldsymbol{U}_1^*(Y_1, Y_2, T; \boldsymbol{\theta}_1) = \boldsymbol{U}_1(Y_1, T; \boldsymbol{\theta}_1) - \sum_{t \in \{0,1\}} \{I(T = t) - \pi_t\} E\{\boldsymbol{U}_1(Y_1, T; \boldsymbol{\theta}_1) \mid Y_2, T = t\}. \qquad (5)$$

We defer to Web Appendix A.1 and Zhang *et al.*, (2008) for more details. From equation (4), it follows that the augmenting term can be re-written in terms of $\mathrm{E}_{1,Y_{2i}} = E(Y_1 \mid Y_{2i}, T = 1)$ and $\mathrm{E}_{0,Y_{2i}} = E(Y_1 \mid Y_{2i}, T = 0)$, which need to be estimated. For binary $Y_2$, they can be estimated from sample averages, and more generally from two distinct logistic regression models, as proposed by Zhang *et al.* (2008). We let $\hat{\mathrm{E}}_{1,Y_{2i}}$ and $\hat{\mathrm{E}}_{0,Y_{2i}}$ denote these estimates and let $\hat{\pi}_1 = \frac{\sum_{i=1}^{n} T_i}{n}$ denote our estimate of $\pi_1$. Therefore, the solution to the augmented estimating equation is

$$\hat{\beta}_1^{\mathrm{Aug}} = \log\left(\frac{\sum_{i=1}^{n}\{T_i Y_{1i} - (T_i - \hat{\pi}_1)\hat{\mathrm{E}}_{1,Y_{2i}}\}}{\sum_{i=1}^{n} T_i} \times \frac{\sum_{i=1}^{n}(1 - T_i)}{\sum_{i=1}^{n}\{(1 - T_i)Y_{1i} + (T_i - \hat{\pi}_1)\hat{\mathrm{E}}_{0,Y_{2i}}\}}\right),$$

with variance estimated from the sandwich formula (see Web Appendix A.2). We refer to this method as the "Augmented" approach (Aug).

In addition to the secondary outcome, baseline covariates $\boldsymbol{W}$ may be used to augment the estimating equation, as in Tsiatis *et al.* (2008) and Zhang *et al.* (2008). In that case, we let $\hat{\beta}_1^{\mathrm{Aug}_{Y_2 \boldsymbol{W}}}$ be part of the solution to the augmented estimating equation based on the slightly modified version of the estimating function given in (5),

$$\boldsymbol{U}_1^*(Y_1, Y_2, \boldsymbol{W}, T; \boldsymbol{\theta}_1) = \boldsymbol{U}_1(Y_1, T; \boldsymbol{\theta}_1) - \sum_{t \in \{0,1\}} \{I(T = t) - \pi_t\} \mathrm{E}\{\boldsymbol{U}_1(Y_1, T; \boldsymbol{\theta}_1) \mid Y_2, \boldsymbol{W}, T = t\}.$$

### 2.3. Reducing bias in observational studies

We now consider estimating $\beta_1$ in an observational study with unmeasured confounders. In the following sub-sections, we present methods to reduce bias from confounding. We note that the validity of these methods relies on the sets of assumptions that are summarized in **TABLE 1**.

**TABLE 1-** Model and assumptions needed to justify estimators of $\beta_1$ that use data on $Y_2$ to reduce bias from unmeasured confounders, $\boldsymbol{A}$. Four scenarios that describe the available measured covariates are presented (column 1), together with their associated estimator (column 2), the model (column 3) and additional assumptions sufficient to yield consistent estimates of $\beta_1$ (column 4).

| Scenario | Estimator | Model | Additional assumptions |
|---|---|---|---|
| 1. No measured covariates | Joint-NC | $\mathrm{E}(Y_1 \mid T, \boldsymbol{A}) = g(\boldsymbol{A}) \exp(\mu_1 + \beta_1 T)$, $E(Y_2 \mid T, \boldsymbol{A}) = h(\boldsymbol{A}) \exp(\mu_2)$. | C2: There is a positive $k > 0$ such that $h(\boldsymbol{a}) = k \times g(\boldsymbol{a})$ for every $\boldsymbol{a}$. We write this $\exists k > 0 / \forall \boldsymbol{a}, h(\boldsymbol{a}) = k \times g(\boldsymbol{a})$. |
| 2. A few large strata indexed by $\boldsymbol{W}$ | SS-Joint (Joint-NC separately by stratum and combined). | $\mathrm{E}(Y_1 \mid T, \boldsymbol{W}, \boldsymbol{A}) = g_{\boldsymbol{W}}(\boldsymbol{A}) \exp(\mu_{1\boldsymbol{W}} + \beta_1 T)$, $\mathrm{E}(Y_2 \mid T, \boldsymbol{W}, \boldsymbol{A}) = h_{\boldsymbol{W}}(\boldsymbol{A}) \exp(\mu_{2\boldsymbol{W}})$. | $C2^*$: $\forall \boldsymbol{w} \in \{\boldsymbol{w}_1, \ldots, \boldsymbol{w}_K\}, \exists k_{\boldsymbol{w}} > 0 / \forall \boldsymbol{a}$, $h_{\boldsymbol{w}}(\boldsymbol{a}) = k_{\boldsymbol{w}} \times g_{\boldsymbol{w}}(\boldsymbol{a})$. |
| Continued | | | |

TABLE 1

Continued

| | | | |
|---|---|---|---|
| 3.<br>Many strata indexed by $\boldsymbol{W}$ | Joint-MH | $\mathrm{E}(Y_1 \mid T, \boldsymbol{W}, \boldsymbol{A}) = g_{\boldsymbol{W}}(\boldsymbol{A}) \exp(\mu_{1\boldsymbol{W}} + \beta_1 T)$,<br>$\mathrm{E}(Y_2 \mid T, \boldsymbol{W}, \boldsymbol{A}) = h_{\boldsymbol{W}}(\boldsymbol{A}) \exp(\mu_{2\boldsymbol{W}})$. | $\mathrm{C2}^*$ and $\mathrm{C3}^*$:<br>$\mathrm{E}\{g_{\boldsymbol{W}}(\boldsymbol{A}) \mid T = 1, W\}/\mathrm{E}\{g_{\boldsymbol{W}}(\boldsymbol{A}) \mid T = 0, \boldsymbol{W}\} = \psi_{\boldsymbol{W}} = \psi$, independent of W.<br>A sufficient condition for $\mathrm{C3}^*$ is<br>$C1^*$: $E\{g_{\boldsymbol{W}}(\boldsymbol{A}) \mid T = t, \boldsymbol{W}\} = exp\{f(t) + l_{\boldsymbol{W}}\}$, with stratum-specific constants $l_{\boldsymbol{W}}$. |
| 4.<br>Continuous and categorical covariates $\boldsymbol{W}$ | Joint-Reg<br>Implied regression equations (12) and (14). | $\mathrm{E}(Y_1 \mid T, \boldsymbol{W}, \boldsymbol{A}) = g(\boldsymbol{A}) \exp\{\mu_1 + \beta_1 T + q(\boldsymbol{W})\}$,<br>$\mathrm{E}(Y_2 \mid T, \boldsymbol{W}, \boldsymbol{A}) = h(\boldsymbol{A}) \exp\{\mu_2 + s(\boldsymbol{W})\}$. | C1:<br>$\mathrm{E}\{g(\boldsymbol{A}) \mid T = t, \boldsymbol{W}\} = \exp\{f(t) + l(\boldsymbol{W})\}$.<br>and C2 and<br>$C3$: $q^*$ and $s^*$ in equations (12) and (14) are correctly specified functions that depend on a finite set of parameters. |

### 2.3.1. Scenario with no measured covariates

In the absence of measured covariates ($\boldsymbol{W} = \emptyset$), equations (1) and (2) reduce to

$$\mathrm{E}(Y_{1i} \mid T_i, \boldsymbol{A}_i) = g(\boldsymbol{A}_i) \exp(\mu_1 + \beta_1 T_i), \tag{6}$$

and

$$\mathrm{E}(Y_{2i} \mid T_i, \boldsymbol{A}_i) = g(\boldsymbol{A}_i) \exp(\mu_2), i \in \{1, \dots, n\}. \tag{7}$$

Because unlike in randomized studies, $T_i$ and $\boldsymbol{A}_i$ are not independent, taking the conditional expectation of $g(\boldsymbol{A}_i)$ given $T_i$ in equation (6) leads to

$$E(Y_{1i} \mid T_i) = \exp(\mu_1^* + \beta_1^* T_i), \tag{8}$$

where $\beta_1^* = \beta_1 + \log\left[\frac{\mathrm{E}\{g(\boldsymbol{A})|T=1\}}{\mathrm{E}\{g(\boldsymbol{A})|T=0\}}\right]$. In other words, $\beta_1^*$ equals $\beta_1$ plus a constant due to the unmeasured confounders, $\boldsymbol{A}$. Therefore, solving the estimating equation based on model (8) will result in a biased estimate of $\beta_1$. However, a similar calculation in equation (7) yields

$$E(Y_{2i} \mid T_i) = \exp(\mu_2^* + \beta_2^* T_i), \tag{9}$$

where $\beta_2^* = \log\left[\frac{\mathrm{E}\{g(\boldsymbol{A})|T=1\}}{\mathrm{E}\{g(\boldsymbol{A})|T=0\}}\right]$. The amount of confounding due to $\boldsymbol{A}$ is thus identical for both outcomes, and $\beta_1 = \beta_1^* - \beta_2^*$. We therefore estimate $\boldsymbol{\theta} \equiv (\mu_1^*, \beta_1^*, \mu_2^*, \beta_2^*)$ from the joint estimating equation $\sum_{i=1}^{n} \boldsymbol{U}(Y_{1i}, Y_{2i}, T_i; \boldsymbol{\theta}) = 0$, with

$$\boldsymbol{U}(Y_1, Y_2, T; \boldsymbol{\theta}) = \begin{pmatrix} \frac{Y_1 - p_1}{1 - p_1} \\ \frac{T(Y_1 - p_1)}{1 - p_1} \\ Y_2 - p_2 \\ T(Y_2 - p_2) \end{pmatrix}, \tag{10}$$

and $p_1 = \exp(\mu_1^* + \beta_1^* T)$ and $p_2 = \exp(\mu_2^* + \beta_2^* T)$. We refer to this approach as the "Joint" approach with no covariates (Joint-NC or JNC). The Joint-NC estimate of the treatment effect on the primary outcome is $\hat{\beta}_1^{\text{JNC}} = \hat{\beta}_1^{*\text{JNC}} - \hat{\beta}_2^{*\text{JNC}} = \log\left\{\frac{\sum_{i=1}^n T_i Y_{1i}}{\sum_{i=1}^n (1-T_i) Y_{1i}} \times \frac{\sum_{i=1}^n (1-T_i) Y_{2i}}{\sum_{i=1}^n T_i Y_{2i}}\right\}$, where $\hat{\beta}_1^{*\text{JNC}}$ and $\hat{\beta}_2^{*\text{JNC}}$ are the estimates from solving the estimating equation with estimation function given in equation (10). Its variance can be obtained from the sandwich formula, by adding the variances of $\hat{\beta}_1^{*\text{JNC}}$ and $\hat{\beta}_2^{*\text{JNC}}$, and subtracting two times their covariance (see Web Appendix B.1). As indicated in **TABLE 1**, the function $g$ in equation (7) can be relaxed to any function $h = k \times g$ for some $k > 0$ without changing $\hat{\beta}_1^{\text{JNC}}$.

#### 2.3.2. Scenario with measured covariates

Suppose we want to estimate $\beta_1$ and have measured potential confounders $\boldsymbol{W}$. We first review standard methods for covariate adjustment that ignore unmeasured confounders before introducing methods that account for them.

##### 2.3.2.1. Standard adjustment for measured confounders only

We consider standard stratification methods and regression methods to adjust for $\boldsymbol{W}$. Let $\boldsymbol{W}$ be a $K$-dimensional stratum indicator vector. $\boldsymbol{W}$ can take one of the $K$ values $\boldsymbol{w}_k$ that have a component 1 in position $k$ and 0 elsewhere. Using Mantel-Haenszel-type weights (Tarone, 1981), we estimate the log-relative risk by

$$\hat{\beta}_1^{\text{MH}} = \log\left(\frac{\sum_{k=1}^K \frac{n_{0k} X_{1k}}{n_k}}{\sum_{k=1}^K \frac{n_{1k} Z_{1k}}{n_k}}\right), \tag{11}$$

with $X_{1k} = \sum_{i=1}^{n} T_i\, Y_{1i}\, 1(\boldsymbol{W}_i = \boldsymbol{w}_k)$, $Z_{1k} = \sum_{i=1}^{n}(1 - T_i)\, Y_{1i}\, 1(\boldsymbol{W}_i = \boldsymbol{w}_k)$, $n_{1k} = \sum_{i=1}^{n} T_i\, 1(\boldsymbol{W}_i = \boldsymbol{w}_k)$, $n_{0k} = \sum_{i=1}^{n}(1 - T_i)\, 1(\boldsymbol{W}_i = \boldsymbol{w}_k)$ and $n_k = \sum_{i=1}^{n} 1(\boldsymbol{W}_i = \boldsymbol{w}_k)$. We call the estimate in equation (11) MH. If relative risks are homogeneous across strata of $\boldsymbol{W}$, MH is asymptotically efficient with a few large strata (Nurminen, 1981) and consistent with sparse strata (Greenland and Robins, 1985). Alternatively, we can regress $Y_1$ on $T$ and $\boldsymbol{W}$ by fitting the model

$$E(Y_{1i} \mid T_i, \boldsymbol{W}_i) = \exp\{\mu_1^* + \beta_1^* T_i + q^*(\boldsymbol{W}_i | \boldsymbol{\alpha}_1^*)\}, \qquad (12)$$

where $q^*$ is a specified parametric function with parameters $\boldsymbol{\alpha}_1^*$. Additional details, including variances, are discussed in Web Appendix B.2.1.

Clearly, these regression and stratification methods do not yield unbiased estimates of $\beta_1$ in the presence of unmeasured confounders. Next, we present modifications based on $Y_2$ that address unmeasured confounders as well as $\boldsymbol{W}$.

#### 2.3.2.2. Adjustment for both measured and unmeasured confounders

We first introduce a modification to the Joint-NC approach that accounts for bias from unmeasured confounders. Suppose $\boldsymbol{W}$ indicates strata and that there are only a few large strata. Within stratum $\boldsymbol{w}$, we can create a Joint-NC estimate $\hat{\beta}_{1\boldsymbol{w}}^{\text{JNC}}$. These stratum-specific estimates can be efficiently combined by taking a weighted combination with weights inversely proportional to $\widehat{\text{Var}}\left(\hat{\beta}_{1\boldsymbol{w}}^{\text{JNC}}\right)$. The only assumption required is that models (8) and (9) hold within each stratum of $\boldsymbol{W}$, as summarized in Scenario 2 of **TABLE 1**. We refer to this method as the "Stratum-Specific Joint" approach (SS-Joint, or SSJ).

Suppose instead there are many, possibly sparse, strata. To account for unmeasured confounders, we can take the logarithm of the ratio of two Mantel-Haenszel-type estimates, one for the primary and one for the secondary outcome, to obtain the Joint-MH (or JMH) estimate given by

$$\hat{\beta}_1^{JMH} = \log\left(\frac{\sum_{k=1}^{K} \frac{n_{0k} X_{1k}}{n_k}}{\sum_{k=1}^{K} \frac{n_{1k} Z_{1k}}{n_k}} \times \frac{\sum_{k=1}^{K} \frac{n_{1k} Z_{2k}}{n_k}}{\sum_{k=1}^{K} \frac{n_{0k} X_{2k}}{n_k}}\right), \tag{13}$$

where $X_{1k} = \sum_{i=1}^{n} T_i\, Y_{1i}\, 1(\boldsymbol{W}_i = \boldsymbol{w}_k)$, $Z_{1k} = \sum_{i=1}^{n} (1 - T_i)\, Y_{1i}\, 1(\boldsymbol{W}_i = \boldsymbol{w}_k)$, $X_{2k} = \sum_{i=1}^{n} T_i\, Y_{2i}\, 1(\boldsymbol{W}_i = \boldsymbol{w}_k)$, $Z_{2k} = \sum_{i=1}^{n} (1 - T_i)\, Y_{2i}\, 1(\boldsymbol{W}_i = \boldsymbol{w}_k)$, $n_{1k} = \sum_{i=1}^{n} T_i\, 1(\boldsymbol{W}_i = \boldsymbol{w}_k)$, $n_{0k} = \sum_{i=1}^{n} (1 - T_i)\, 1(\boldsymbol{W}_i = \boldsymbol{w}_k)$ and $n_k = \sum_{i=1}^{n} 1(\boldsymbol{W}_i = \boldsymbol{w}_k)$. The estimate $\hat{\beta}_1^{\mathrm{JMH}}$ is consistent under conditions $\mathtt{C1}^*$ and $\mathtt{C2}^*$ in **TABLE 1**.

To account for unmeasured confounders with regression models, we introduce Joint-Reg (or JReg) by fitting both equation (12) and a second regression model based on equation (14):

$$E(Y_{2i} \mid T_i, \boldsymbol{W}_i) = \exp\{\mu_2^* + \beta_2^* T_i + s^*(\boldsymbol{W}_i | \boldsymbol{\alpha}_2^*)\},\ i \in \{1, \dots, n\}. \tag{14}$$

Assuming models (12) and (14), we let $\widehat{\boldsymbol{\theta}}^{*\mathrm{JReg}}$ be the value of $\boldsymbol{\theta}^* \equiv (\mu_1^*, \beta_1^*, \boldsymbol{\alpha}_1^*, \mu_2^*, \beta_2^*, \boldsymbol{\alpha}_2^*)$ that solves the joint estimating equations and then use the estimate $\hat{\beta}_1^{\mathrm{JReg}} = \hat{\beta}_1^{*\mathrm{JReg}} - \hat{\beta}_2^{*\mathrm{JReg}}$. Sufficient conditions for $\hat{\beta}_1^{\mathrm{JReg}}$ to be consistent for $\beta_1$ are in Scenario 4 of **TABLE 1**. See Web Appendix B.2.2 for additional details, including variance calculations and motivation for the Joint-MH and Joint-Reg estimators.

As indicated in **TABLE 1**, increasingly stringent assumptions are required to justify estimates of $\beta_1$ that take measured covariates $\boldsymbol{W}$ and unmeasured confounders $\boldsymbol{A}$ into account, as one proceeds from adjustment on a few large strata (Scenario 2) to many possibly sparce strata (Scenario 3) to full regression analysis (Scenario 4). We note that the simulations (SECTION 3) use a realistic data generation mechanism where these assumptions are not met. Therefore, it is not anticipated that estimates like $\hat{\beta}_1^{\text{JReg}}$ and $\hat{\beta}_1^{\text{JMH}}$ will remove all confounding bias.

### 2.4. Composite outcome

Many vaccine trials focus on the effect of the vaccine on the composite endpoint of having an HPV 16 and/or HPV 18 infection. Letting $Y_{1i}^{(j)}$ be the binary indicator of an infection with the $j^{\text{th}}$ targeted type for subject $i$, with $j \in \{16,18\}$, we denote the composite primary endpoint by $Y_{1i}^{C} = Y_{1i}^{(16)} + Y_{1i}^{(18)} - Y_{1i}^{(16)}Y_{1i}^{(18)}$. Suppose equation (1) holds separately for to $Y_{1i}^{(16)}$ and $Y_{1i}^{(18)}$, namely

$$\mathrm{E}\left(Y_{1i}^{(j)} \mid T_i, \boldsymbol{W}_i, \boldsymbol{A}_i\right) = g(\boldsymbol{A}_i)\exp\left\{\mu_1^{(j)} + \beta_1^{(j)} T_i + q^{(j)}(\boldsymbol{W}_i)\right\}, j \in \{16,18\}.$$

Web Appendix C.1 defines the log relative risk, $\beta_1^C$, for $Y_1^C$ in terms of these parameters. In our simulations in SECTION 3.3, we analyze $Y_1^C$ to reflect conventional analysis of the composite endpoint, which yields the estimates $\hat{\beta}_1^{\text{C,JMH}}$ and $\hat{\beta}_1^{\text{C,JReg}}$. Web Appendix C.2 calculates the correlation between $Y_1^C$ and $Y_2$ under the assumption that each $Y_2^{(j)}$ follows equation (2) with a common $g(\boldsymbol{A})$ but its own intercept $\mu_2^{(j)}$ and function $s^{(j)}$. The estimates $\hat{\beta}_1^{\text{C,JMH}}$ and $\hat{\beta}_1^{\text{C,JReg}}$ are biased for $\beta_1^{\text{C}}$ in observational studies because equation (1) is not satisfied for $Y_1^C$. In Web Appendix C.3, we present sufficient conditions so that model (1) holds approximately for $Y_1^C$. For comparison, we performed similar simulations with a single targeted type as primary outcome, for which equation (1) is satisfied, to assess how much bias in $\hat{\beta}_1^{\text{C,JMH}}$ and $\hat{\beta}_1^{\text{C,JReg}}$ is due to misspecification of model (1) (Web Appendix E.3.3).

In contrast, in randomized studies, a model of the form (3) holds for $Y_1^C$, with $\beta_1^C$ replacing $\beta_1$ and we can use method Aug to estimate $\beta_1^C$ with increased precision.

## 3. SIMULATIONS

### 3.1. Simulation methods

We evaluate how well the methods in SECTIONS 2.2 and 2.3 can estimate $\beta_1^C$, the log relative risk, for the targeted HPV types in simulated randomized and observational studies. Vaccine efficacy is related to $\beta_1^C$ through $VE = 1 - \exp(\beta_1^C)$. We considered multiple scenarios, defined by the models described below and the parameter values listed in **WEB TABLES 1-6**. Overall, the parameters were chosen so that some simulated studies have similar incidence and prevalence of infections as in the original Costa Rica Vaccine Trial (CVT, Herrero *et al.*, 2008). We note that in the simulated observational studies, which we tried to make realistic, condition C1 in **TABLE 1** does not hold.

We simulated studies with $n \in \{5 \times 10^3, 10^4, 5 \times 10^4\}$ individuals. We simulated two observed covariates, $\boldsymbol{W} = (W^{\text{site}}, W^{\text{age}})$, which represent an individual's geographic location and age. $W^{\text{site}}$ can take the values $w_{\text{site}} \in \Omega_{W^{\text{site}}} = \{0,1,2\}$, each with probability 1/3, and $W^{\text{age}}$ can take values $w_{\text{age}} \in \Omega_{W^{\text{age}}} = \{15, 15.5, \dots, 20.5, 21\}$, with equal probability; we let $\Omega_{\boldsymbol{W}} = \Omega_{W^{\text{site}}} \times \Omega_{W^{\text{age}}}$. We simulated one unobserved covariate, $A$, which is a numeric value reflecting a woman's relative risk of acquiring an infection. $A$ can take the values $a \in \Omega_A = \{a_{\text{low}} = 0, a_{\text{medium}}, a_{\text{high}}\}$ with probabilities $\{p_{a_{\text{low}}|\boldsymbol{W}}, p_{a_{\text{medium}}|\boldsymbol{W}}, p_{a_{\text{high}}|\boldsymbol{W}}\}$; possible values and probabilities are defined in **WEB TABLES 1-3**. For randomized trials, we simulated a treatment indicator $T \sim \mathcal{B}(0.5)$, where $\mathcal{B}(p)$ is a Bernoulli variable with probability $p$. For observational studies, $T \mid A, \boldsymbol{W} \sim \mathcal{B}(p_{T|A,\boldsymbol{W}})$, where $p_{T|A,\boldsymbol{W}}$ was chosen so that the overall probability of receiving the vaccine is 0.5 and is given by

$$\frac{1}{2}\times\frac{\exp\left(\gamma_T+\delta_T W^{\text{age}}+\eta_T W^{\text{site}}+\vartheta_T A\right)}{\left(1+\exp\left(\gamma_T+\delta_T W^{\text{age}}+\eta_T W^{\text{site}}+\vartheta_T A\right)\right)}\times\left\{\sum_{\boldsymbol{w}\in\Omega_{\boldsymbol{W}},a\in\Omega_A}\frac{\exp\left(\gamma_T+\delta_T w^{\text{age}}+\eta_T w^{\text{site}}+\vartheta_T a\right)}{1+\exp\left(\gamma_T+\delta_T w^{\text{age}}+\eta_T w^{\text{site}}+\vartheta_T a\right)}\frac{1}{3\times 13}p_{a|\boldsymbol{w}}\right\}^{-1}.$$

We simulated infection by HPV 16 and HPV 18, $\left(Y_1^{(16)},Y_1^{(18)}\right)$, and we simulated infection by the $N_{\text{NT}}$ = 20 non-targeted HPV types, $\left(Y_2^{(1)},\dots,Y_2^{(N_{\text{NT}})}\right)$, from which we computed the primary and secondary outcomes $Y_1^C=Y_1^{(16)}+Y_1^{(18)}-Y_1^{(16)}Y_1^{(18)}$ and $Y_2=\sum_{j\in J_{\text{NT}}}Y_2^{(j)}$. We assumed $Y_1^{(j)}\mid T,A,\boldsymbol{W}\sim\mathcal{B}\left(A\times p_1^{(j)}\right)$ with $p_1^{(j)}=\exp\left\{\mu_1^{(j)}+\beta_1^{(j)}T+\alpha_1^{(j)}W^{\text{age}}+\sum_{w\in\Omega_{W^{\text{site}}}}\lambda_w^{(j)}I\left(W^{\text{site}}=w\right)\right\}$, $j\in\{16{,}18\}$. We chose $\mu_1^{(16)}$and $\mu_1^{(18)}$ so that $\left(\mathrm{P}\left(Y_1^{(16)}=1\right),\mathrm{P}\left(Y_1^{(18)}=1\right)\right)\in\{(0.032,0.015),(0.05,0.05),(0.14,0.07)\}$. Similarly, $Y_2^{(j)}\mid T,A,\boldsymbol{W}\sim\mathcal{B}\left(A\times p_2^{(j)}\right)$ with $p_2^{(j)}=\exp\left\{\mu_2^{(j)}+\alpha_2^{(j)}W^{\text{age}}+\sum_{w\in\Omega_{W^{\text{site}}}}\mu_w^{(j)}I\left(W^{\text{site}}=w\right)\right\}$, $j\in\{1,\cdots,N_{\text{NT}}\}$.

We estimated $\beta_1^C$ in randomized clinical trials using Aug and UnAug (SECTION 2.2). For comparison, we also estimated $\beta_1^C$ when the primary outcome estimating equation is augmented with a function of baseline covariates $\boldsymbol{W}$ only (method "Aug$_{\boldsymbol{W}}$", as proposed by Zhang *et al.* (2008)), and when the estimating equation is augmented with a function of both $\boldsymbol{W}$ and $Y_2$ (method "Aug$_{Y_2\boldsymbol{W}}$"). Then we estimated $\beta_1^C$ in observational studies using MH, Joint-NC, Joint-MH and Joint-Reg (SECTION 2.3). For each scenario, we simulated 10,000 datasets.

### 3.2. Results for the randomized trials

The results of the simulations for several scenarios when $n=10{,}000$ are presented in **TABLE 2**. All methods estimate the effect of $T$ on $Y_1^C$ without bias (results not shown), and all yield coverages of confidence intervals near 0.95. Variance ratios compare the variance of UnAug to that of the other methods. Variance ratios larger than one indicate improved efficiency from the alternative method. Method Aug leads to an estimate with smaller variance than that of UnAug (column 1 of **TABLE 2**) and of $\text{Aug}_{\boldsymbol{W}}$ (column 2 of **TABLE 2**). Using information on both baseline covariates and non-targeted strains reduces the variance further (column 3 of **TABLE 2**). The efficiency gain from Aug, compared to UnAug, measured by the variance ratio, ranges from 1.127 to 1.011, depending on the incidence of targeted strains and other factors. The efficiency gain from $\text{Aug}_{Y_2\boldsymbol{W}}$ ranges from 1.176 to 1.021. Plots of actual variances for the various methods are shown in **WEB FIGURE 3**.

We note that when the correlation between $Y_1^C$ and $Y_2$ is small, there is minimal benefit to using Aug (or $\text{Aug}_{Y_2\boldsymbol{W}}$). The correlation tends to be smaller when the probabilities of infection are lower.

### 3.3. Results for observational studies

We compare estimates from adjusting only on measured covariates (MH), using only the non-targeted strains (Joint-NC), from using stratification on measured confounders and using non-targeted strains (Joint-MH), and regression on measured confounders with correction using non-targeted strains (Joint-Reg) (see **FIGURE 2** and **TABLE 3**). Recall that when using method Joint-Reg, parametric forms need to be specified for $q^*$and $s^*$ in model (12) and model (14). Misspecification of $q^*$and $s^*$ may affect the performance of Joint-Reg, and as noted in Web Appendix B.2, $q^*$ and $s^*$ usually differ from $q$ and $s$, respectively. Here, in order to be flexible, we used $q^*\left(w^{\text{age}}, w^{\text{site}}\right) = \alpha_1^{(j)*} w^{\text{age}} + \tilde{\alpha}_1^{(j)*} w^{\text{age}2} + \sum_{w \in \Omega_{W^{\text{site}}}} \lambda_w^{(j)*} I\left(w^{\text{site}} = w\right)$, for $\left(w^{\text{age}}, w^{\text{site}}\right) \in \Omega_{\boldsymbol{W}}$, and $s^*$ with similar form.

| Variance ratio for Aug | Variance ratio for $\text{Aug}_{\boldsymbol{W}}$ | Variance ratio for $\text{Aug}_{Y_2\boldsymbol{W}}$ | Coverage of 95% CI on $\beta_1^C$ for UnAug | Coverage of 95% CI on $\beta_1^C$ for Aug | Coverage of 95% CI on $\beta_1^C$ for $\text{Aug}_{\boldsymbol{W}}$ | Coverage of 95% CI on $\beta_1^C$ for $\text{Aug}_{Y_2\boldsymbol{W}}$ | Incidence probabilities $\left(P\left(Y_1^{(16)}=1\right), P\left(Y_1^{(18)}=1\right)\right)$ | Values of the unmeasured covariate ($a_{\text{low}}$, $a_{\text{medium}}$, $a_{\text{high}}$) | Correlation between $Y_1^C$ and $Y_2$ |
|---|---|---|---|---|---|---|---|---|---|
| 1.117 | 1.04 | 1.164 | 0.949 | 0.952 | 0.951 | 0.951 | (0.14, 0.07) | (0, 1, 2.5) | 0.376 |
| 1.088 | 1.039 | 1.133 | 0.952 | 0.951 | 0.951 | 0.949 | (0.14, 0.07) | (0, 1, 2) | 0.339 |
| 1.127 | 1.039 | 1.176 | 0.952 | 0.952 | 0.952 | 0.95 | (0.14, 0.07) | (0, 0.75, 2) | 0.388 |
| 1.048 | 1.018 | 1.068 | 0.949 | 0.95 | 0.951 | 0.951 | (0.05, 0.05) | (0, 1, 2.5) | 0.259 |
| 1.034 | 1.018 | 1.053 | 0.951 | 0.953 | 0.953 | 0.954 | (0.05, 0.05) | (0, 1, 2) | 0.231 |
| 1.053 | 1.018 | 1.073 | 0.952 | 0.952 | 0.952 | 0.953 | (0.05, 0.05) | (0, 0.75, 2) | 0.268 |
| 1.019 | 1.009 | 1.029 | 0.951 | 0.952 | 0.951 | 0.952 | (0.032, 0.015) | (0, 1, 2.5) | 0.179 |
| 1.011 | 1.009 | 1.021 | 0.953 | 0.951 | 0.953 | 0.952 | (0.032, 0.015) | (0, 1, 2) | 0.158 |
| 1.02 | 1.009 | 1.032 | 0.949 | 0.95 | 0.951 | 0.95 | (0.032, 0.015) | (0, 0.75, 2) | 0.185 |

**TABLE 2-** Confidence interval (CI) coverages for $\beta_1^C$ and the ratio of the empirical variance of $\hat{\beta}_1^{C,\text{UnAug}}$ to that of $\hat{\beta}_1^{C,\text{Aug}}$, $\hat{\beta}_1^{C,\text{Aug}_{\boldsymbol{W}}}$ and $\hat{\beta}_1^{C,\text{Aug}_{Y_2\boldsymbol{W}}}$ for 10,000 simulated randomized trials with $n = 10{,}000$ participants. The variance ratio, a measure of efficiency gain from Aug, $\text{Aug}_{\boldsymbol{W}}$ or $\text{Aug}_{Y_2\boldsymbol{W}}$ (compared to UnAug) varies across the various simulated scenarios. Incidence probabilities refer to the incidence of HPV types 16 and 18 in the entire study population of vaccinated (50 %) and unvaccinated (50%) participants.

The analysis that ignores unmeasured confounders (MH) yields severely upwardly biased estimates of $\beta_1^C$. Joint-NC reduces bias somewhat, but the bias is still large (**FIGURE 2**). Both Joint-MH and Joint-Reg reduce bias almost to zero over a range of scenarios. Joint-MH consistently has less bias than Joint-Reg, however. The findings for Joint-MH and Joint-Reg suggest that these procedures are robust, because they worked quite well (but not perfectly) even though condition C1 in **TABLE 1** was violated and even though the analysis used a composite HPV 16/18 outcome. In Web Appendix E.3.3, we present additional simulations that show that the small bias for Joint-MH and Joint-Reg is mostly due to residual unmeasured confounding (probably from violation of condition C1 in **TABLE 1**) rather than from misspecification of equation (1) for the composite endpoint.

## 3.4. Sensitivity of analyses to the assumption that vaccine has no effect on targeted outcomes

As described in Web Appendix E.3.1, we conducted sensitivity analyses in which the vaccine has an effect on the non-targeted outcomes, contrary to our assumption (see equation (2)). We used estimates of $\beta_2^{(j)}$ from incident and prevalent infections in the CVT (parameters $\boldsymbol{\nu}_1$), more extreme estimates based on confidence limits ($\boldsymbol{\nu}_2$), and an extreme set of $\beta_2^{(j)}$ all with the same negative value ($\boldsymbol{\nu}_3$).

For $\boldsymbol{\nu}_1$ in randomized trials, the estimate $\hat{\beta}_1^{\text{Aug}}$ is nearly unbiased and more efficient than $\hat{\beta}_1^{\text{UnAug}}$, measured by mean square error (**WEB FIGURE 4** and **WEB TABLE 7**). For $\boldsymbol{\nu}_2$, there is a small bias and loss of confidence interval coverage with $\hat{\beta}_1^{\text{Aug}}$ (**WEB FIGURE 5** and **WEB TABLE 8**).

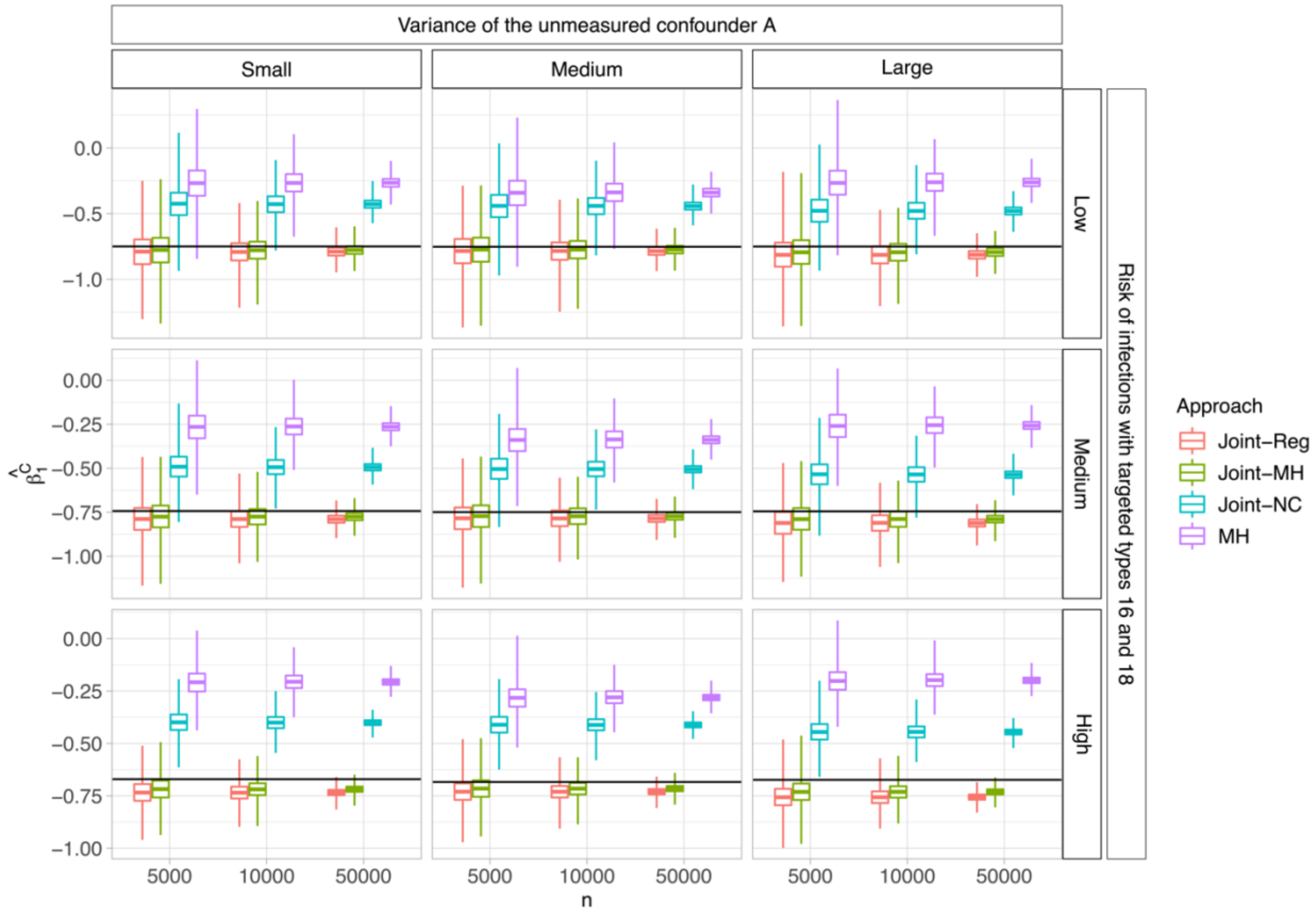

**FIGURE 2 -** Box plots of estimates of $\beta_1^C$ from MH (purple), Joint-NC (blue), Joint-MH (green) and Joint-Reg (red) from 10,000 simulated observational studies of size $n = 5{,}000, 10{,}000$ or $50{,}000$. The nine scenarios cover a variety of settings. The rows represent Low ($\mathrm{P}\left(Y_1^{(16)} = 1\right) = 0.032, \mathrm{P}\left(Y_1^{(18)} = 1\right) = 0.015$), Medium ($\mathrm{P}\left(Y_1^{(16)} = 1\right) = 0.05, \mathrm{P}\left(Y_1^{(18)} = 1\right) = 0.05$), and High ($\mathrm{P}\left(Y_1^{(16)} = 1\right) = 0.14, \mathrm{P}\left(Y_1^{(18)} = 1\right) = 0.07$) risk of infections with the targeted types. The columns represent Small ($a_{\text{low}} = 0$, $a_{\text{medium}} = 0.75$, $a_{\text{high}} = 2$), Medium ($a_{\text{low}} = 0$, $a_{\text{medium}} = 1$, $a_{\text{high}} = 2$), and Large ($a_{\text{low}} = 0$, $a_{\text{medium}} = 1$, $a_{\text{high}} = 2.5$) variance of the unmeasured covariate, $A$. The mean of $Y_2$ is 1.75. The true value of $\beta_1^C$ is indicated by the horizontal black line and is calculated as in Web Appendix C.1.

| Bias with MH | Bias with Joint-NC | Bias with Joint-MH | Bias with Joint-Reg | Incidence probabilities $\left(P\left(Y_1^{(16)}=1\right), P\left(Y_1^{(18)}=1\right)\right)$ | Values of unmeasured confounder ($a_{low}$, $a_{medium}$, $a_{high}$) | Correlation between $Y_1^{C}$ and $Y_2$ |
|---|---|---|---|---|---|---|
| 0.476 | 0.228 | -0.058 | -0.083 | (0.14, 0.07) | (0, 1, 2.5) | 0.359 |
| 0.404 | 0.273 | -0.032 | -0.046 | (0.14, 0.07) | (0, 1, 2) | 0.328 |
| 0.465 | 0.27 | -0.048 | -0.064 | (0.14, 0.07) | (0, 0.75, 2) | 0.371 |
| 0.489 | 0.209 | -0.044 | -0.066 | (0.05, 0.05) | (0, 1, 2.5) | 0.244 |
| 0.412 | 0.245 | -0.024 | -0.036 | (0.05, 0.05) | (0, 1, 2) | 0.221 |
| 0.48 | 0.249 | -0.032 | -0.046 | (0.05, 0.05) | (0, 0.75, 2) | 0.252 |
| 0.49 | 0.272 | -0.043 | -0.063 | (0.032, 0.015) | (0, 1, 2.5) | 0.168 |
| 0.414 | 0.31 | -0.022 | -0.033 | (0.032, 0.015) | (0, 1, 2) | 0.152 |
| 0.484 | 0.322 | -0.028 | -0.041 | (0.032, 0.015) | (0, 0.75, 2) | 0.174 |

**TABLE 3-** Bias in estimates of $\beta_1^{C}$, for $\hat{\beta}_1^{C,MH}$, $\hat{\beta}_1^{C,JNC}$ $\hat{\beta}_1^{C,JMH}$ and $\hat{\beta}_1^{C,JReg}$ based on 10,000 simulated observational studies with $n=10{,}000$ participants. Incidence probabilities refer to the incidence of HPV types 16 and 18 in the entire study population of vaccinated (50 %) and unvaccinated (50%) participants.

For $\boldsymbol{\nu}_3$, there is appreciable bias and loss of coverage of confidence intervals (**WEB FIGURE 6** and **WEB TABLE 9**), but even in this extreme case the bias is small in comparison with $\beta_1^C$.

For $\boldsymbol{\nu}_1$, $\boldsymbol{\nu}_2$ and $\boldsymbol{\nu}_3$ in observational studies, methods Joint-MH and Joint-Reg still reduce bias substantially (**WEB FIGURES 7-9** and **WEB TABLES 10-12**). Therefore, reduction in confounding bias from using Joint-MH and Joint-Reg is robust to violations of the assumption of no effect of vaccination on non-targeted strains.

## 4. EXAMPLES BASED ON VACCINE DATA

We consider a randomized trial and an observational study of HPV vaccine effects. We first consider the Costa Rica Vaccine Trial (CVT). The CVT was one of the original randomized trials that demonstrated that three doses of Cervarix ® prevents HPV 16/18 infections and subsequent pre-cancerous lesions. We then consider the ongoing India Study. The India study was initially designed to compare a two and three dose regimen of quadrivalent Gardasil ®, but due to the suspension of the trial during the vaccination stage, observational data from the trial are currently being used to compare infection rates in an unvaccinated cohort and in a cohort who received 1-dose of the vaccine.

### 4.1. Increasing efficiency in the Costa Rica Vaccine Trial (CVT)

We estimated $\beta_1^C$, the effect of the HPV vaccine on the cumulative incidence of infection by HPV 16 and/or HPV 18, from the CVT (Herrero *et al.*, 2008). Here $Y_1^C$ was the primary outcome, and $Y_2$ was the total number of infections with $N_{\mathrm{NT}}$ = 20 non-targeted HPV types (for which the vaccine had not shown statistically significant effects, types 6, 11, 34, 35, 39, 40, 42, 43, 44, 51, 52, 53, 54, 56, 58, 59, 66, 70, 74 and 68/73). In these data, we have access to information on age at baseline and province for the 7168 participating women, and 49.9% of them received three doses of the vaccine. The correlation between $Y_1^C$ and $Y_2$ was 0.283 overall, and 0.376 in the unvaccinated group, and the cumulative incidences of HPV 16 and 18 in the entire study population were 0.076 and 0.049, respectively.

We obtained $\hat{\beta}_1^{\mathrm{C,UnAug}} =$ -1.323 (95% CI = [ -1.484, -1.161]) and $\hat{\beta}_1^{\mathrm{C,Aug}} =$ -1.305 (95% CI = [-1.462, -1.148]). Corresponding estimates of vaccine efficacy, $1 - \exp(\hat{\beta}_1^{\mathrm{C}})$, are 0.734 (95% CI = [0.687, 0.773]) and 0.729 (95% CI = [0.683, 0.768]). For comparison, we also estimated $\beta_1^C$ with methods $\mathrm{Aug}_{\boldsymbol{W}}$ and $\mathrm{Aug}_{Y_2\boldsymbol{W}}$, where $\boldsymbol{W}$ includes age at baseline and province. We obtained $\hat{\beta}_1^{\mathrm{C,Aug}\boldsymbol{W}}$ = -1.32 (95% CI = [ -1.482, -1.158]) and $\hat{\beta}_1^{\mathrm{C,Aug}\boldsymbol{W}Y_2}$ = -1.304 (95% CI = [-1.461, -1.147]), with corresponding estimates of vaccine efficacy, $1 - \exp(\hat{\beta}_1^{\mathrm{C}})$, 0.733 (95% CI = [0.686, 0.773]) and 0.729 (95% CI = [0.682, 0.768]). From randomization, all estimates are unbiased, but using augmented estimating equations reduced the variance for the estimates. The ratio of the variance of $\hat{\beta}_1^{\mathrm{C,UnAug}}$ to that of $\hat{\beta}_1^{\mathrm{C,Aug}}$ was 1.058, indicating an efficiency gain of 5.8% from Aug, whereas the efficiency gain from $\mathrm{Aug}_{\boldsymbol{W}}$, compared to UnAug, was only of 0.1%. Using $\mathrm{Aug}_{Y_2\boldsymbol{W}}$ yielded an efficiency gain of 5.9%.

Vaccine efficacy is usually assessed by using incident persistent infections that have smaller probabilities than cumulative infections. We found vaccine efficacies against persistent infections based on $\hat{\beta}_1^{\mathrm{C,UnAug}}$ and $\hat{\beta}_1^{\mathrm{C,Aug}}$ to be near 0.85, which is higher than for cumulative infections that include transient non-persistent infections. The efficacy 0.85 is slightly lower than the value 0.947 reported for incident persistent infections from two large trials (Tota *et al.*, 2020), possibly because we used an intent-to-treat analysis. Because the rates for incident persistent infections were very low, the gain in statistical efficiency from using Aug was smaller than for cumulative incidence (see Web Appendix F.1).

### 4.2. Analysis of observational data from India

We estimated the effect, $\beta_1^C$, of the HPV vaccine on cumulative incident infection by HPV 16 and/or HPV 18 from an observational study in India (Sankaranarayanan *et al.*, 2018). We used $Y_1^C$ as the primary outcome, and $Y_2$, the total number of infections with $N_{\mathrm{NT}}$ = 17 non-targeted HPV types (types 26, 31, 33, 35, 39, 45, 51, 52, 53, 56, 58, 59, 66, 68, 70, 73 and 82), as the secondary outcome. Among the $n = 4098$ women evaluated, 63.8% received one dose of the vaccine, and the remainder were unvaccinated. The correlation between primary and secondary outcome was 0.255 overall, and 0.349 in the unvaccinated group, and the cumulative incidences of targeted HPV strains 16 and 18 were 0.038 and 0.02, respectively. Measured baseline covariates included geographic site, age and education level.

Unadjusted analyses (UnAug) indicated a strong protective effect of vaccine, $\hat{\beta}_1^{\mathrm{C,UnAug}}$ = -1.02 (**TABLE 4**) and $\widehat{\mathrm{VE}}^{\mathrm{UnAug}}$ = 0.639. Using MH, we found that adjustment for age alone or age $\times$ education had little effect, but adjustment for age $\times$ site reduced the estimate to -0.621, with corresponding estimate of vaccine efficacy 0.462. All methods that used $Y_2$ to account for unmeasured confounding, whether they adjusted for measured covariates (Joint-MH or Joint-Reg) or not (Joint-NC) further reduced the estimate of $\beta_1^C$ to the range -0.569 to -0.429, leading to an estimate of the VE ranging from 0.349 to 0.434 (not shown). For example, with stratification adjustment on age $\times$ site, the estimate of $\beta_2$ is -0.621-(-0.565) = -0.056, indicating that unmeasured confounding decreases apparent risk, $\hat{\beta}_1^{\mathrm{C,MH}}$, in vaccinated subjects. This could happen if those who were not vaccinated engaged in riskier sexual behavior, for example.

We performed similar analyses for incident persistent infections (Web Appendix F.2). The estimated vaccine efficacies were significantly higher, ranging from 0.962 to 0.989. However, as there is little information to estimate $\beta_1^C$ when using that endpoint, we obtained broad confidence intervals.

## 5. DISCUSSION

We proposed methods for using HPV infection outcomes in strains not targeted by a vaccine to increase precision of estimates of vaccine effects on targeted strains in randomized trials and to reduce bias in estimated effects on targeted strains from unmeasured confounders in observational studies. We found only modest increases in precision in HPV vaccine trials, but in other clinical trial settings, where the primary outcome probability is larger, the improvements in precision from using secondary outcomes may be larger. We believe our paper is the first to suggest a valid method for using post-treatment negative control events to improve precision in randomized trials.

We found important bias reductions in observational HPV studies from using non-targeted strains. This approach may be useful for many observational HPV vaccine studies, because key covariates like sexual activity are usually not recorded, whereas outcomes for non-targeted strains are usually available but not analyzed.

Our results depend on several key assumptions. First, we assume that the treatment has no effect on non-targeted outcomes ("negative control outcomes"). This assumption needs to be checked empirically. Data from randomized studies can be used to estimate the vaccine effects on non-targeted strains (or all non-targeted strains combined) and their confidence intervals. For example, data from the CVT in SECTION 4.1 yielded an estimate of the vaccine effect on the total number of non-targeted infections of -0.02 (95% CI = [-0.076, 0.036]); see also Table 2 in Tota et *al.* (2020). Because such data does not prove that there is no effect, but only that the effect is likely to be small ("evidence of lack of appreciable effect"), it is necessary to see how sensitive our methods are to small violations of the assumption of no effect. Our sensitivity analyses for randomized studies indicated a detectable bias toward the null, but the bias was small in comparison with the vaccine effect on targeted strains. In observational studies, sensitivity analyses indicated that our procedures for reducing bias from unmeasured confounders were hardly perturbed by violations of the assumption that vaccination had no effect on non-targeted strains. In other applications, where the negative control outcome assumption is less certain, more extensive sensitivity analyses would be required. Otherwise, the use of non-targeted strains to increase efficiency of analyses of randomized trials is robust to modeling assumptions, violations of which can decrease the gains in efficiency but not bias the trial results.

| Method | Observed covariates | Estimated vaccine effect, $\hat{\beta}_1^C$ | 95% CI on vaccine effect |
|---|---|---|---|
| UnAug | - | -1.02 | [-1.281, -0.759] |
| MH | Age (11 strata) | -0.999 | [-1.216, -0.782] |
| MH | Age (5 categories) × Site (9 categories) | -0.621 | [-1.004, -0.237] |
| MH | Age (5 categories) × Education (5 categories) | -1.032 | [-1.262, -0.803] |
| Joint-NC | - | -0.429 | [-0.697, -0.162] |
| Joint-MH | Age (11 strata) | -0.477 | [-0.833, -0.121] |
| Joint-Reg | Age (continuous) | -0.447 | [-0.736, -0.157] |
| Joint-MH | Age (5 categories) × Site (9 categories) | -0.565 | [-0.882, -0.248] |
| Joint-Reg | Age (continuous), Site (9 categories) | -0.555 | [-0.952, -0.158] |
| Joint-MH | Age (5 categories) × Education (5 categories) | -0.569 | [-0.906, -0.231] |
| Joint-Reg | Age (continuous), Education (5 categories) | -0.468 | [-0.758, -0.179] |

**TABLE 4-** Estimates of $\beta_1^C$ for cumulative incident HPV 16 or 18 infections from the observational study in India (Sankaranarayanan et *al.*, 2018). Analysis with UnAug does not adjust for measured or unmeasured confounders, analyses labeled MH adjust for measured confounders only, analyses labeled Joint-MH and Joint-Reg adjust for measured confounders and for unmeasured confounding via the non-targeted outcome data, $Y_2$, and analysis labeled Joint-NC adjusts for unmeasured confounding only.

More assumptions are needed for the analysis of observational data using non-targeted strains. One is that unmeasured confounders affect targeted and non-targeted strains proportionally, namely that the effects of the unmeasured confounders on the targeted and non-targeted strains have the same shape function $g$, but not necessarily the same magnitude. This is plausible for unmeasured sexual activity, for example. If there are only a few large strata, proportionality is only required within strata (**TABLE 1**). Another is that the ratio of expected multiplier effects, $\mathrm{E}\{\mathrm{g}(\boldsymbol{A}) \mid T=1, \boldsymbol{W}=\boldsymbol{w}\}/\mathrm{E}\{\mathrm{g}(\boldsymbol{A}) \mid T=0, \boldsymbol{W}=\boldsymbol{w}\}$, is independent of $\boldsymbol{W}$. Condition C1 in **TABLE 1** is sufficient to guarantee this independence. Despite violations of the latter assumption in realistic simulations, procedures based on these assumptions drastically reduced bias from unmeasured confounding.

We proposed two approaches for incorporating measured confounders and non-targeted outcomes to eliminate confounding in observational studies. One approach defined strata in terms of measured confounders. If there are only a few large strata, few assumptions are needed (Scenario 2, **TABLE 1**). If there are many possibly sparse strata, an additional homogeneity condition is required (Scenario 3, **TABLE 1**). A regression approach can handle continuous covariates, at the risk of stronger modeling assumptions. In particular, in addition to the proportionality and homogeneity assumptions, regression requires correct specification of the parametric functions in equations (12) and (14) (Scenario 4, **TABLE 1**). In simulations we found that stratification yielded slightly less bias than regression methods, possibly because stratification approaches require less stringent assumptions. Both methods performed vastly better than methods that did not use non-targeted outcomes, however. In fact, the performance of both approaches was surprisingly good in simulations, especially as conditions like C1 were not satisfied.

We were surprised that methods motivated by analysis of a single HPV type performed so well for a composite endpoint. Web Appendix C.3 gives conditions under which equation (1) holds approximately for a composite endpoint. Provided infection with the component HPV types is rare, equation (1) holds approximately if either the $\beta_1^{(j)}$ or $q^{(j)}$ are homogeneous. In our reported and unreported simulations these conditions were not satisfied, yet biases when estimating $\beta_1^{\mathrm{C}}$ were small for Joint-MH and Joint-Reg. Simulations reported in Web Appendices E.3.2 and E.3.3 suggest that the small residual bias is mainly from failure to control for unmeasured confounding, not from misspecification by equation (1). Thus, at least for rare outcomes like $Y_1^{(16)}$ and $Y_1^{(18)}$ we can recommend our procedures for composite endpoints. However, if the effects of the vaccine differ vastly among the targeted types, we also suggest using our methods to analyze and report the effect of the vaccine against each type separately.

The expected value in equation (1) is loglinear in the effects of $T$, $q(\boldsymbol{W})$ and $\log\{g(\boldsymbol{A})\}$. In randomized trials in which treatment is independent of measured and unmeasured confounders, the treatment parameter $\beta_1$ is the same whether the analysis is conditional on $\boldsymbol{W}$ and $\boldsymbol{A}$ or as a population effect averaged over measured and unmeasured covariates (Gail, Wieand and Piantadosi, 1984). This model also approximates a logistic regression for rare outcomes. If logistic outcomes are not rare, the methods we presented for randomized trials remain valid for estimating the population average parameter. The loglinear form (or a linear model) is needed for our methods for observational data, however.

**ACKNOWLEDGMENTS**

We would like to thank Dr. Richard Muwonge, Dr. Partha Basu, and the Indian HPV vaccine study group for graciously sharing the data from their HPV trial.

This work utilized the computational resources of the NIH HPC Biowulf cluster. (http://hpc.nih.gov)

**DATA AVAILABILITY STATEMENT**

R code and functions used for the simulations in SECTION 3 are available in the Supporting Information of this article and on GitHub at https://github.com/Etievant/NonTargetedHPV.

Methods Joint-NC, SS-Joint and Joint-MH in SECTIONS 1 and 2.3.2.2 are also available on CRAN in the NegativeControlOutcomeAdjustment R package (https://cran.r-project.org/web/packages/NegativeControlOutcomeAdjustment/index.html).

We are not authorized to release the clinical data used in SECTION 4.

## SUPPORTING INFORMATION

Web Appendices, Tables and Figures referenced in SECTIONS 2.2, 2.3, 2.4, 3.1, 3.2, 3.3, 3.4, 4.1 and 4.2 are available with this paper at the Biometrics website on Wiley Online Library.

R code and functions used for the simulations in Web Appendix E are available in the Supporting Information of this article and on GitHub at https://github.com/Etievant/NonTargetedHPV.

We are not authorized to release the clinical data used in Web Appendix F.